\documentstyle[preprint,aps,epsf,floats]{revtex}

\begin{document}

\preprint{\tighten \vbox{\hbox{} }}

\title{Bounds on Heavy-to-Heavy Mesonic Form Factors}
\author{Cheng-Wei Chiang\footnote{\tt chengwei@andrew.cmu.edu} and Adam
K. Leibovich\footnote{\tt adaml@cmuhep2.phys.cmu.edu}}

\address{
Department of Physics,
Carnegie Mellon University,
Pittsburgh, PA 15213}

\maketitle

{\tighten
\begin{abstract}
We provide upper and lower bounds on the form factors for $B
\rightarrow D, D^*$ by utilizing inclusive heavy quark effective
theory sum rules.  These bounds are calculated to leading order in
$\Lambda_{\rm QCD}/m_Q$ and $\alpha_s$.  The $O(\alpha_s^2\beta_0)$
corrections to the bounds at zero recoil are also presented.  We
compare our bounds with some of the form factor models used in the
literature.  All the models we investigated failed to fall within the
bounds for the combination of form factors $(\omega^2 -
1)/(4\omega)|\omega h_{A2}+h_{A3}|^2$.
\end{abstract}
}


\newpage

\section{Introduction}


Currently, heavy-to-heavy form factors are mainly derived from models.
Since models are used in Monte Carlos, which in turn feeds into
studies of backgrounds and efficiencies in many experimental settings,
constraining them is very important.  Model independent information on
the form factors can be obtained from the Heavy Quark Effective Theory
(HQET).  HQET \cite{IW89,IW90,FGGW90} vastly simplifies the
nonperturbative calculation of form factors by relating all of them
in the infinite mass limit to a single universal Isgur-Wise function,
$\xi$, which describes the nonperturbative physics of the light
degrees of freedom in the heavy mesons.  This function is normalized
to unity at zero recoil, where the heavy meson in the final state has
the same velocity as the initial one.  Nonetheless, HQET does not
predict the explicit form of the Isgur-Wise function.

Since it is not possible to calculate heavy-to-heavy form factors
from first principles, the next best thing to do is
to theoretically bound them.\footnote{See, however, \cite{BGL95} for
model independent parameterizations of the form factors.}  A set of
model independent bounds on form factors have been derived
\cite{BSUV95,BR97}, and can serve as a consistency condition for
phenomenological models.  Bounds to ${\mathcal O} (1/m_Q)$ at
arbitrary momentum transfer had been presented in \cite{BR97,BR98} for
heavy-to-heavy and heavy-to-light form factors.  However, as noted in
\cite{BR97,BR98,BLRW97}, the lower bound is sensitive to the
perturbative corrections, and leading order in $\alpha_s$ correction
should be included to provide more reliable bounds.  Therefore, in
this work we present an analysis of the leading QCD corrections on the
bounds of individual heavy-to-heavy, more specifically $B \rightarrow
D^{(*)}\ell\bar\nu$, form factors.  The information of these bounds
will help us rule out unrealistic models, and may in turn be applied
to certain decay amplitudes that are of interest.

This paper is organized as follows.  In Section II, we review the
derivation of the sum rules used to obtain the model-independent
bounds on form factors defined in Section III.  Section III also gives
the proper combinations of structure functions used later for the
bounds on each of the form factors.  In Section IV, we perform the
tree level nonperturbative expansion and first order QCD perturbative
calculation to eliminate the leading uncertainty in the bounds.
Section V provides the bounds on individual form factors explicitly
with the structure functions given in the Appendix, and discusses
various influences of the parameters appearing in the expansion of the
bounds.  Some popular form factor models are compared with our bounds
in Section VI.  Order $\alpha_S^2 \beta_0$ corrections to the bounds
at zero recoil are computed in Section VII.  The conclusion of our
work is summarized in Section VIII.  The Appendix lists the
perturbative corrections to the structure functions.


\section{Review of sum rules}

The sum rules are derived by relating the inclusive decay rate,
calculated using the operator product expansion (OPE) and perturbative
QCD, to the sum of exclusive decay rates.  In the following, we follow
\cite{BR97,BLRW97} in the derivation of the bounds.

First, consider the time ordered product of two currents between $B$
mesons in the momentum space,
\begin{eqnarray}
\label{structurefn}
T^{\mu\nu}& = & \frac{i}{2M_{B}} \int d^4x \, e^{-iq \cdot x} 
   \left< B(v) \right| T(J^{\mu\dagger}(x)J^{\nu}(0)) \left| B(v) \right> 
   \nonumber \\
& = &-g^{\mu\nu}T_1 + v^{\mu}v^{\nu}T_2 
   + i \epsilon^{\mu\nu\alpha\beta}q_{\alpha}v_{\beta}T_3 
   + q^{\mu}q^{\nu}T_4 + (q^{\mu}v^{\nu}+v^{\mu}q^{\nu})T_5,
\end{eqnarray}
where $J^\mu$ is a $b\to c$ axial or vector current.  The time ordered
product can be expressed as a sum over hadronic or partonic
intermediate states.  The sum over hadronic states includes the matrix
element $\langle H|J|B\rangle$, where $H=D,D^{(*)}$ is the final state
heavy meson of interest.  After inserting a complete set of states and
dotting with four vectors $a_{\mu}^*a_{\nu}$, one obtains
\begin{eqnarray}\label{tmnexp}
T(\epsilon) &=& \frac1{2M_B}\sum_X (2\pi)^3\delta^3(\vec{p}_X+\vec q) 
  \frac{\left|\left<X\right|a\cdot J\left|B\right>\right|^2}{E_X-E_H-\epsilon}
  \nonumber\\ 
&& + \frac1{2M_B}\sum_X (2\pi)^3\delta^3(\vec{p}_X-\vec q) 
  \frac{\left|\left<B\right|a\cdot J\left|X\right>\right|^2}
  {\epsilon+E_X+E_H-2M_B},
\end{eqnarray}
where $T(\epsilon) \equiv a_{\mu}^*T^{\mu\nu}a_{\nu}$,
$\epsilon=M_B-E_H-v \cdot q$, and the sum over $X$ includes the usual
$d^3p/2E_X$ for each particle in the state $X$.  We choose to work in
the rest frame of the $B$ meson, $p = M_B v$, and the $z$ axis
pointing in the direction of $\vec q$.  We will hold $q_3$ fixed while
analytically continuing $v \cdot q$ to the complex plane.  $E_H =
\sqrt{M_H^2+q_3^2}$ is the $H$ meson energy.  There are two cuts in
the complex $\epsilon$ plane, $0<\epsilon<\infty$, corresponding to
the decay process $b\to c$, and $-\infty<\epsilon<-2E_H$,
corresponding to two $b$ quarks and a $\bar c$ quark in the final
state.  The second cut will not be important for our discussion.

The integral over $\epsilon$ of the time ordered product,
$T(\epsilon)$, times a weight function, $\epsilon^n
W_\Delta(\epsilon)$ can be computed perturbatively in QCD
\cite{BR97,BLRW97}.  For simplicity, we pick the weight function
$W_\Delta(\epsilon) = \theta(\Delta-\epsilon)$, which corresponds to
summing over all hadronic resonances up to the excitation energy
$\Delta$ with equal weight.  Relating the integral with the hard
cutoff to the exclusive states requires local duality at the scale
$\Delta$.  $\Delta$ must be chosen large enough so that the structure
functions can be calculated perturbatively.

Taking the zeroth moment of $T(\epsilon)$
\begin{eqnarray}
\label{zerothmmnt}
\frac1{2\pi i}\int_{C}d\epsilon\,\theta(\Delta-\epsilon)\,T(\epsilon) &=&
\sum_X \theta(E_X - E_M - \Delta) (2 \pi)^3 \delta^3(\vec{q} + \vec{p}_X)
\frac{\left | \langle X|J\cdot a | B\rangle \right |^2}{2 M_B}\nonumber\\
&\geq&
   \frac{\left| \left< H(v') \right| a \cdot J \left|
      B(v) \right> \right|^2}{4M_B E_H}
\end{eqnarray}
gives an upper bound on the matrix element.

The first moment of $T(\epsilon)$ gives
\begin{eqnarray}
\label{firstmmnt}
&&\frac1{2\pi i}\int_C d\epsilon\,\epsilon\,\theta(\Delta-\epsilon)T(\epsilon)
\nonumber\\
&&\phantom{\frac1{2\pi i}\int_C d\epsilon}
 =\sum_{X\not=H} \theta(\Delta-E_X+E_H)\,(E_X-E_H)\,(2\pi)^3 \delta^3(p_X+q)
 \frac{\left|\left<X\right|a\cdot J\left|B\right>\right|^2}{4M_B E_X}
 \nonumber \\
&&\phantom{\frac1{2\pi i}\int_C d\epsilon}
\geq
 (E_1-E_H) \sum_{X\not=H} \theta(\Delta-E_X+E_H)(2\pi)^3 \delta^3(p_X+q)
 \frac{\left|\left<X\right|a\cdot J\left|B\right>\right|^2}{4M_B E_X}.
\end{eqnarray}
where $E_1$ denotes the first excited state that is more massive than
$H$ meson.  Here the validity of the inequality relies on the
assumption that multiparticle final states with less energy than $E_1$
contribute negligibly.  This assumption is true in large $N_c$,
and is also confirmed by current experimental data.

A lower bound can be formed by combining Eqs.~(\ref{zerothmmnt}) and
(\ref{firstmmnt}) to be
\begin{equation}
\label{lowb}
\frac{\left| \left< H(v') \right| a \cdot J \left| B(v) \right> 
   \right|^2}{4M_B E_H}
   \geq \frac{1}{2\pi i} \int_{C} d\epsilon \, \theta(\Delta-\epsilon) \, 
   T(\epsilon) \left( 1-\frac{\epsilon}{E_1-E_H} 
   \right). \nonumber
\end{equation}
Therefore, we find the bounds
\begin{eqnarray}
\label{uplowb}
\frac{1}{2\pi i} \int_{C} d\epsilon \, \theta(\Delta-\epsilon) \, T(\epsilon)
   \left( 1-\frac{\epsilon}{E_1-E_H} \right)
&\leq& \frac{\left| \left< H(v') \right| a \cdot J \left| B(v) \right>
   \right|^2}{4M_B E_H} \nonumber \\
&\leq& \; \frac{1}{2\pi i} \int_{C} d\epsilon \, \theta(\Delta-\epsilon) \,
   T(\epsilon).
\end{eqnarray}
As emphasized in \cite{BLRW97}, the upper bound is essentially model
independent while the lower bound relies on the assumption about the
final state spectrum.  These bounds can be used for the decays at
arbitrary momentum transfer $q^2$ and are good for both heavy mesons
and baryons.  (For baryons, a spin sum
$\frac{m_H}{2J+1}\sum_{S,S^{\prime}}$ needs to be included in front of the bounded factor.)

Since $1/(E_1-E_H) \sim 1/\Lambda_{\rm QCD}$, the lower bounds will be
limited to order in $1/m_Q$.  The upper bounds, on the other hand, can
be calculated to order $1/m_Q^2$ without additional HQET parameters.


\section{Hadronic side}

The hadronic matrix elements for semi-leptonic decay of a $B$ meson
into a pseudoscalar meson $P$ or a vector meson $V$ may be
parameterized as
\begin{eqnarray}
\label{param}
\frac{\left< P(v^\prime)\mid V^\mu \mid B(v)\right>}{\sqrt{M_P M_B}} & = &
   h_+(\omega) \left( v + v^\prime \right)^\mu
   + h_-(\omega)  \left( v-v^\prime \right)^\mu, \nonumber \\ 
\frac{\left< V(v^\prime,\varepsilon)\mid V^\mu\mid B(p)\right>}
  {\sqrt{M_V M_B}} &=& i h_V(\omega) \epsilon^{\mu\nu\alpha\beta}
   \varepsilon^*_\nu v^\prime_\alpha v_\beta,  \\
\frac{\left< V(v^\prime,\varepsilon)\mid A^\mu\mid B(v)\right>}{\sqrt{M_VM_B}}
   &=& h_{A_1}(\omega) \, (\omega +1) \varepsilon^{*\mu}
     - \left[ h_{A_2}(\omega) v^\mu + h_{A_3}(\omega) v^{\prime\mu} \right]
     v \cdot \varepsilon^* . 
\nonumber
\end{eqnarray}
$v'$ is the velocity of the final state meson, and the variable
$\omega = v \cdot v^{\prime}$ is a measure of the recoil.  One may
relate $\omega$ to the momentum transfer $q^2$ by $\omega =
(M_B^2+M_H^2-q^2)/(2M_B M_H)$.  Therefore, with a proper choices of
the current $J^{\mu}$ and the four vector $a^{\mu}$, one may readily
select the form factor of interest and the corresponding sum rule for
bounding.  In the heavy quark symmetry limit, these form factors
satisfy relations \cite{IW89,IW90}
\begin{eqnarray}
\label{hqetformfactors}
h_+(\omega) &=& h_V(\omega)=h_{A_1}(\omega)=h_{A_3}(\omega)=\xi(\omega),
   \nonumber \\ 
h_+(\omega) &=& h_{A_1}(\omega) = 0.
\end{eqnarray}

To bound $h_{A_1}$ (or $h_V$), one may choose $J^{\mu}=A^{\mu}$ (or
$V^{\mu}$) and $a^{\mu}=(0,1,0,0)$.  Then the factor to be bounded is
$\frac{(\omega+1)^2}{4\omega} \left| h_{A_1}(\omega) \right|^2 \;
\left( \frac{\omega^2-1}{4\omega} \left| h_V(\omega) \right|^2
\right)$ and the sum rule used to bound is $T_{1OPE}=T_{1Hadronic}$.
The corresponding first excited state more massive than $D^*$ that
contributes to the sum rule is the $J^P=1^+$ state, {\it i.e.} $D_1$,
since scalars do not contribute to $T_1$.

For $h_+$, one may take $J^{\mu}=V^{\mu}$ and $a^{\mu}=(1+
E_D/M_D,0,0,-q_3/M_D)$.  Then the factor to be bounded is
$\frac{(\omega+1)^2}{\omega} \left| h_+(\omega) \right|^2$, and the
combination of structure functions used in bounds is
\begin{eqnarray}
T(\epsilon) &=& -2 \left( 1+\omega \right)T_1+\left(1+\omega\right)^2 T_2 
+ \left( M_B-M_D-\epsilon \right)^2 \left( 1+\omega \right)^2 T_4 \nonumber \\
&&+ 2 \left(M_B-M_D-\epsilon\right)\left(1+\omega \right)^2 T_5.
\end{eqnarray}
Since $a^{\mu} = v^\mu + v^{\prime\mu} $, the first excited resonance
that can contribute in this case is $D_1$, due to the
$\epsilon^{\mu\nu\alpha\beta}$ structure of the $D^*$ form factor.

Similarly, a convenient choice to isolate $h_-$ is to choose
$J^{\mu}=V^{\mu}$ and $a^{\mu}=(1-E_D/M_D,0,0,q_3/M_D)$.  Thus,
\begin{eqnarray}
T(\epsilon) &=&
  - 2 \left( 1-\omega \right) T_1 + \left( 1-\omega \right)^2 T_2
  + \left(M_B+M_D-\epsilon \right)^2 \left(1-\omega \right)^2 T_4\nonumber\\
&&+ 2 \left(M_B+M_D-\epsilon \right) \left( 1-\omega \right)^2 T_5
\end{eqnarray}
bounds $\frac{(\omega-1)^2}{\omega} \left| h_-(\omega)\right|^2$.
$D_1$ would be the first excited resonance for the same reason as in
the case of $h_+$.

It is impossible to single out $h_{A_2}$ and $h_{A_3}$ individually by
any choice of $a^{\mu}$.  One good choice is to take $J^{\mu}=A^{\mu}$
and $a^{\mu}=(E_D/M_D,0,0,-q_3/M_D)$.  Then
\begin{eqnarray}
T(\epsilon) &=&
  -T_1 + \omega^2 T_2 + \left( M_B \omega - M_D^* -\epsilon \right)^2 T_4
  \nonumber \\
&&
  + 2 \omega \left( M_B \omega - M_D^* -\epsilon \right) T_5
\end{eqnarray}
is the combination for bounding $\frac{\omega^2-1}{4\omega} \left|
\omega h_{A_2}(\omega) + h_{A_3}(\omega) \right|^2$.  The first
excited resonance that would contribute in this case would be the
unobserved $D_0^*$.


\section{Partonic side}

Due to the heavy quark masses appearing in the problem, both the
strong coupling constant $\alpha_s(m_Q)$ ($\sim 0.3$ at $2 {\rm\
GeV}$) and $\Lambda_{\rm QCD}/m_Q$ will be good expansion parameters.
Since $\omega$ is never very far from one, expanding in $\omega-1$ is
also possible.  We will keep terms up to order $\alpha_s (w-1)$,
dropping terms of order $\alpha_s (w-1)^2$, $\alpha_s^2$,
$\Lambda_{\rm QCD}^2/m_Q^2$ and $\alpha_s \Lambda_{\rm QCD}/m_Q$.

The structure functions can be decomposed as
\begin{equation}
T_i^{full} = T_i^{1/m} + T_i^{\alpha_s},
\end{equation}
where the term $T_i^{1/m}$ contains the tree and $\Lambda_{\rm
QCD}/m_Q$ contribution to the structure function, which has previously
been calculated \cite{BKSV94,BGM96,MW94}.  For these
nonperturbative corrections, we keep the full $\omega$ dependence.

The leading order $\alpha_s$ corrections consist of bremsstrahlung
radiation of a gluon from the heavy quarks and one loop virtual
corrections.  We expand to first order in $w-1$, defining
\begin{equation}
\label{uvdef}
T_i^{\alpha_s} = \frac{\alpha_s}\pi\,[U_i + (w-1)\,V_i].
\end{equation}
The final results for the functions $U_i$ and $V_i$ are presented in
the Appendix.  The perturbative corrections to $T_1$ were previously
calculated in \cite{KLWG96,BLRW97}, and agree with the results found
here.

We define the moments of the structure functions as
\begin{equation}\label{SFmoments}
\frac1{2\pi i}\int d\epsilon\,\epsilon^n\,T_i =  
I^{(n)}_i + A^{(n)}_i,
\end{equation}
where
\begin{eqnarray}
I^{(n)}_i &=& \frac1{2\pi i}\int d\epsilon\,\epsilon^n\,T_i^{1/m},\nonumber\\
A^{(n)}_i &=& \frac1{2\pi i}\int d\epsilon\,\epsilon^n\,
   \frac{\alpha_s}\pi\,[U_i + (w-1)\,V_i].
\end{eqnarray}
The moments of $T_i^{1/m}$ can be found in \cite{BSUV95,BR97}.  The
moments for the perturbative corrections can be straightforwardly
obtained from the functions in the Appendix.  One thing to be noted is
that the integration variable $\epsilon$ in the bounds was defined in
terms of hadronic variables.  So when relating to the partonic
computations, the corresponding integration variable should be changed
to the one defined by partonic variables, namely, $\epsilon_p =
m_b-m_c-v \cdot q$.  The relation between them is $\epsilon_h =
\epsilon_p + \delta$, with $\delta=E_c-E_H+M_B-m_b$ and $E_c$ being
the energy of the c-quark.  We can now use these definitions to
calculate the bounds on the form factors.


\section{Bounds on individual form factors}

To form the bounds, one just takes the proper moments of the structure
functions to form the combination required in the sum rules given in
Section III.  Corrections of order $\Lambda_{\rm QCD}^2/m_Q^2$,
$\alpha_s^2$, $\alpha_s\,\Lambda_{\rm QCD}/m_Q$, and
$\alpha_s\,(\omega -1)^2$ should be small and have been neglected.  To
this order, the $\Lambda_{\rm QCD}/m_Q$ corrections will depend on 3
HQET parameters; $\bar\Lambda$, $\lambda_1$ and $\lambda_2$.  From the
measured mass difference, $M_{D^*}-M_D$, a very accurate value of
$\lambda_2$ can be determined, $\lambda_2=0.12{\rm\ GeV}^2$.  It is
very difficult to measure the parameters $\bar\Lambda$ and $\lambda_1$
individually, but a certain linear combination is much better
determined \cite{GKLW96}.  We will use three different parameter sets
to show the dependence on $\bar\Lambda$ and $\lambda_1$: (A)
$\bar\Lambda = 0.4 {\rm\ GeV}$ and $\lambda_1 = -0.2{\rm\ GeV}^2$, (B)
$\bar\Lambda = 0.3 {\rm\ GeV}$ and $\lambda_1 = -0.1{\rm\ GeV}^2$, (C)
$\bar\Lambda = 0.5 {\rm\ GeV}$ and $\lambda_1 = -0.3{\rm\ GeV}^2$.

The sum rule for bounding $(\omega+1)^2 \left| h_{A_1}(\omega)
\right|^2/(4\omega)$ uses $T_1$ with the axial-axial current.  The
upper bound is simply the zeroth moment of $T_1$, which is by
Eq.~(\ref{uplowb})
\begin{equation}
\label{hA1upper}
\frac{(\omega+1)^2}{4\omega} \left| h_{A_1}(\omega) \right|^2 \leq
   I_1^{(0)AA} + A_1^{(0)AA}.
\end{equation}
The variable $\lambda$ used for regularizing infrared divergences in
the kinematic region away from zero recoil disappears in the final
result.  The first moment of $T_1$ is needed for the lower bound,
which is
\begin{equation}
\frac{(\omega+1)^2}{4\omega} \left| h_{A_1}(\omega) \right|^2 \geq
   I_1^{(0)AA} + A_1^{(0)AA}
- \frac{1}{E_{D_1}-E_D*}
     \left( I_1^{(1)AA} + A_1^{(1)AA}\right)
\end{equation}
The upper and lower bounds are shown in Fig.~1.\footnote{For all the
figures in this section we take $m_b = 4.8 {\rm\ GeV}$, $m_c = 1.4
{\rm\ GeV}$, $\alpha_s = 0.3$ (corresponding to a scale of about $2
{\rm\ GeV}$), and $\lambda_2 = 0.12 {\rm\ GeV^2}$.  The values of
$\bar\Lambda$ and $\lambda_1$ are discussed in the text.}
\begin{figure}[t]
\centerline{\epsfysize=11truecm  \epsfbox{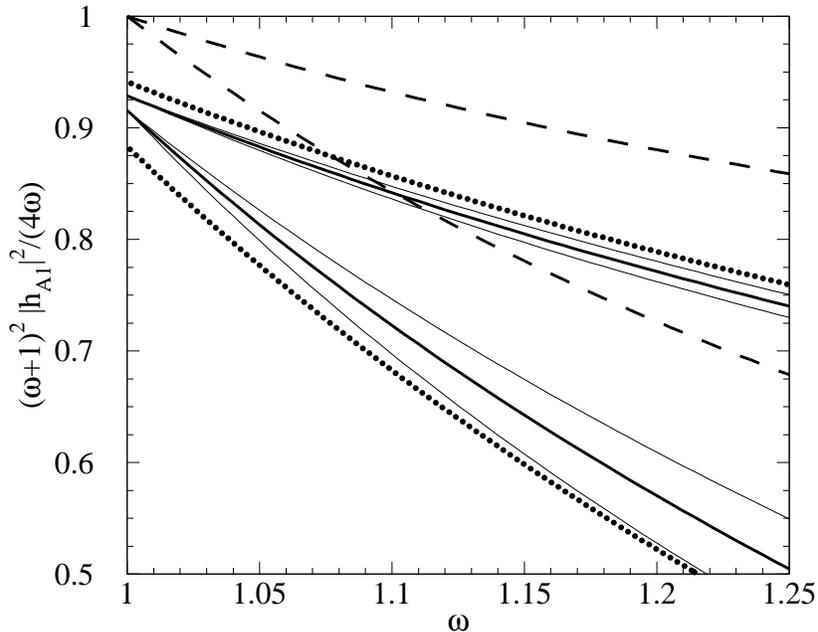} }
\tighten{
\caption[]{\it Upper and lower bounds on $(\omega+1)^2 \left|
h_{A_1}(\omega) \right|^2/(4\omega)$.  The thick solid (dotted) curves
are the upper and lower bounds including perturbative corrections for
HQET parameter set (A) described in the text, and $\Delta = 1{\rm\ GeV}\
(2{\rm\ GeV})$.  The dashed curves are the bounds without perturbative
corrections, also for HQET parameter set (A).  The thin solid curves
show the dependence on $\bar\Lambda$ and $\lambda_1$, using parameter
sets (B) and (C), with $\Delta = 1{\rm\ GeV}$.}}
\end{figure}
For this section, the dotted curves are the bounds without
perturbative corrections using set (A) above, while the solid and
dashed curves represent the bounds including the perturbative
corrections with $\Delta=1 {\rm\ GeV}$ and $\Delta=2 {\rm\ GeV}$,
respectively.  We have shown the bounds in the kinematic range $1 \leq
\omega \lesssim 1.25$, where the higher order correction
$\alpha_s(\omega -1)^2$ should be negligible.  The thin solid curves
use the other HQET parameter sets (B) and (C).

A similar set of bounds for $(\omega^2-1) \left| h_V(\omega)
\right|^2/(4\omega)$ can be obtained readily by changing $AA$
(axial-axial currents) to $VV$ (vector-vector currents) in the above
formulae.  The bounds in this case are shown in Fig.~2.  
\begin{figure}[t]
\centerline{\epsfysize=11truecm  \epsfbox{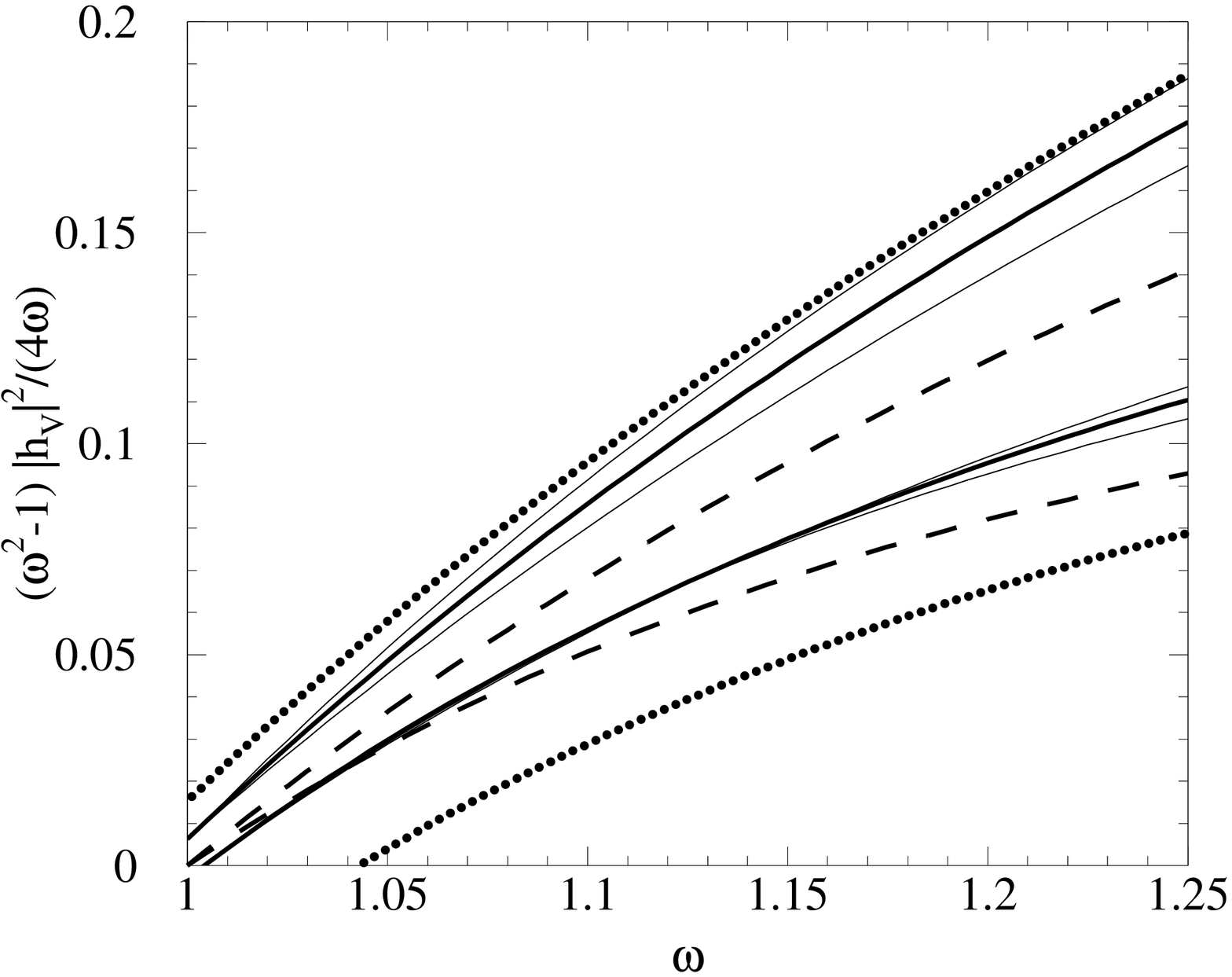} }
\tighten{
\caption[]{\it Upper and lower bounds on $(\omega^2-1) \left|
h_V(\omega) \right|^2/(4\omega)$.  The thick solid (dotted) curves are
the bounds with perturbative corrections for HQET parameter set (A), and
$\Delta = 1{\rm\ GeV}\ (2{\rm\ GeV})$.  The dashed curves are the
bounds without perturbative corrections, also for HQET parameter set
(A).  The thin solid curves show the dependence on $\bar\Lambda$ and
$\lambda_1$, using parameter sets (B) and (C), with $\Delta = 1{\rm\
GeV}$.}}
\end{figure}
The tree level bounds go to zero at zero recoil because of the
$\omega-1$ factor.

The bounds on the other form factors involve higher moments of $T_4$
and $T_5$.  The upper bound for $(\omega+1)^2 \left|
h_+(\omega) \right|^2/\omega \leq h_+^{upper}$, where
\begin{eqnarray}
\label{hpupbound}
h_+^{upper}
&=& ( 1+\omega )^2\left\{-2 \frac1{ 1+\omega}I_1^{(0)VV}
 +  I_2^{(0)VV} 
 + I_4^{(2)VV} - 2 (M_B-M_D) I_4^{(1)VV}\right.\\
&& \phantom{( 1+\omega )^2[]} + (M_B-M_D)^2 I_4^{(0)VV}
 + 2 (M_B-M_D) I_5^{(0)VV} - 2 I_5^{(1)VV}
\nonumber \\
&&\phantom{( 1+\omega )^2[]}
-2 \frac1{1+\omega}A_1^{(0)VV}
 + A_2^{(0)VV} 
 + A_4^{(2)VV} - 2 (M_B-M_D) A_4^{(1)VV} \nonumber\\
&&\phantom{( 1+\omega )^2[]}
 + \left. (M_B-M_D)^2 A_4^{(0)VV}
 + 2 (M_B-M_D) A_5^{(0)VV} -
 2 A_5^{(1)VV}\right\}.\nonumber
\end{eqnarray}
The lower bound is $(\omega+1)^2 \left|
h_+(\omega) \right|^2/\omega \geq h_+^{lower}$, where
\begin{eqnarray}
\label{hplowbound}
h_+^{lower} &=& h_+^{upper} \\
&&- \frac{(1+\omega)^2}{M_{D_1} - M_D}\left\{
-2 \frac1{ 1+\omega}I_1^{(1)VV}
 +  I_2^{(1)VV} 
 + I_4^{(3)VV} - 2 (M_B-M_D) I_4^{(2)VV}\right.\nonumber\\
&& \phantom{- \frac{(1+\omega)^2}{M_{D_1} - M_D}[]} + (M_B-M_D)^2 I_4^{(1)VV}
 + 2 (M_B-M_D) I_5^{(1)VV} - 2 I_5^{(2)VV}
\nonumber \\
&&\phantom{- \frac{(1+\omega)^2}{M_{D_1} - M_D}[]}
-2 \frac1{1+\omega}A_1^{(1)VV}
 + A_2^{(1)VV} 
 + A_4^{(3)VV} - 2 (M_B-M_D) A_4^{(2)VV} \nonumber\\
&&\phantom{- \frac{(1+\omega)^2}{M_{D_1} - M_D}[]}
 + \left. (M_B-M_D)^2 A_4^{(1)VV}
 + 2 (M_B-M_D) A_5^{(1)VV} -
 2 A_5^{(2)VV}\right\}.\nonumber
\end{eqnarray}
They are plotted in Fig.~3.  
\begin{figure}[t]
\centerline{\epsfysize=11truecm  \epsfbox{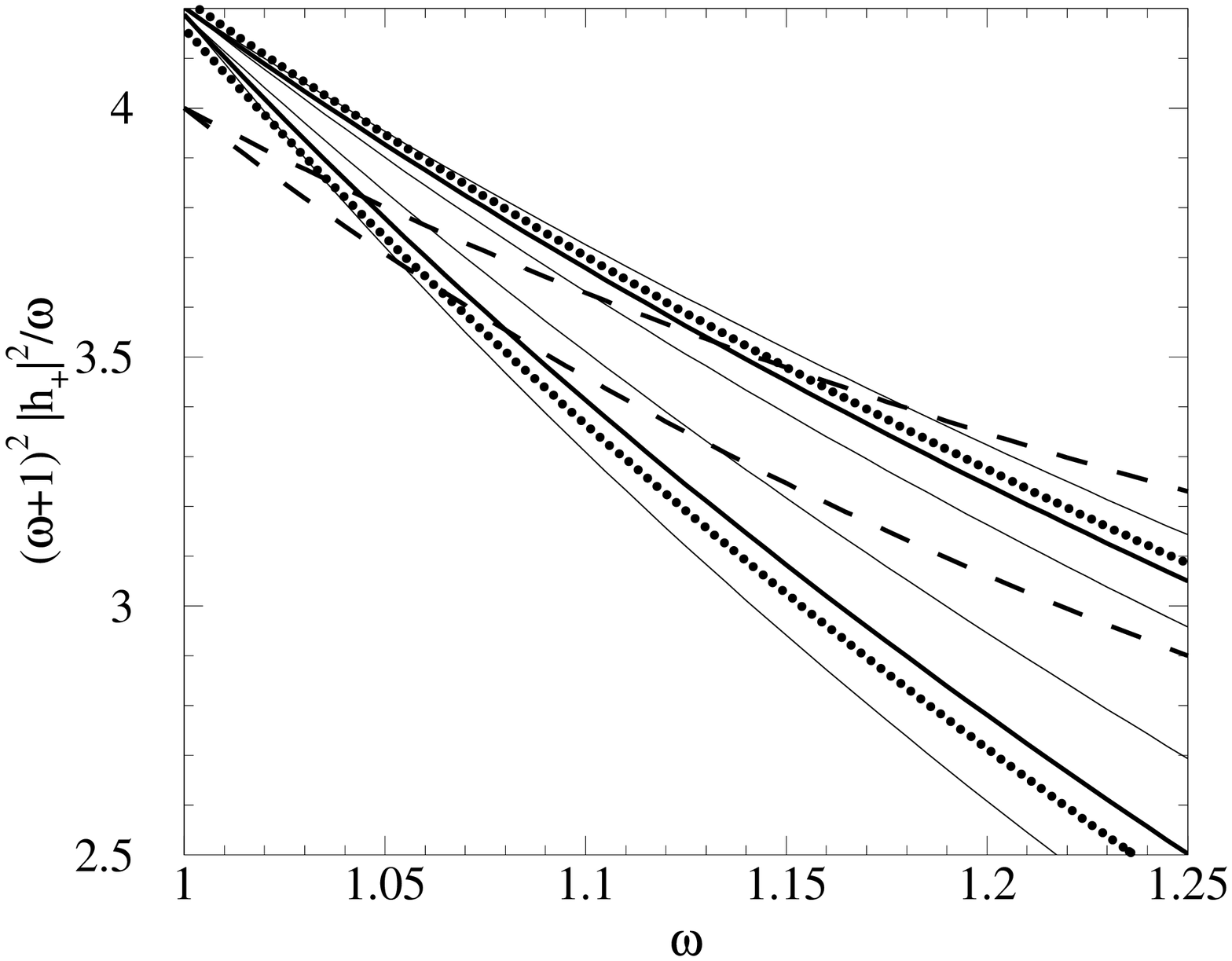} }
\tighten{
\caption[]{\it Upper and lower bounds on $(\omega+1)^2 \left|
h_+(\omega) \right|^2/\omega$ The thick solid (dotted) curves are the
bounds with perturbative corrections for HQET parameter set (A), and
$\Delta = 1{\rm\ GeV}\ (2{\rm\ GeV})$.  The dashed curves are the
bounds without perturbative corrections, also for HQET parameter set
(A).  The thin solid curves show the dependence on $\bar\Lambda$ and
$\lambda_1$, using parameter sets (B) and (C), with $\Delta = 1{\rm\
GeV}$.}}
\end{figure}
The perturbative physics pushes the bounds up from the tree level
bounds in the region near the maximal momentum transfer while drags
them down at large recoil.  Changing $\Delta$ from $1{\rm\ GeV}$ to
$2{\rm\ GeV}$ only slightly loosens both bounds.

Without explicitly writing out the bounds on $(\omega-1)^2
|h_-(\omega)|^2/\omega$, we simply state that they can be obtained
from Eq.~(\ref{hpupbound}) and (\ref{hplowbound}) by: 1) replacing
each $(1 + \omega)$ factor by $(1-\omega)$ and 2) substituting
$M_B-M_D$ by $M_B+M_D$.  The bounds on this form factor are shown in
Fig.~4.  
\begin{figure}[t]
\centerline{\epsfysize=11truecm  \epsfbox{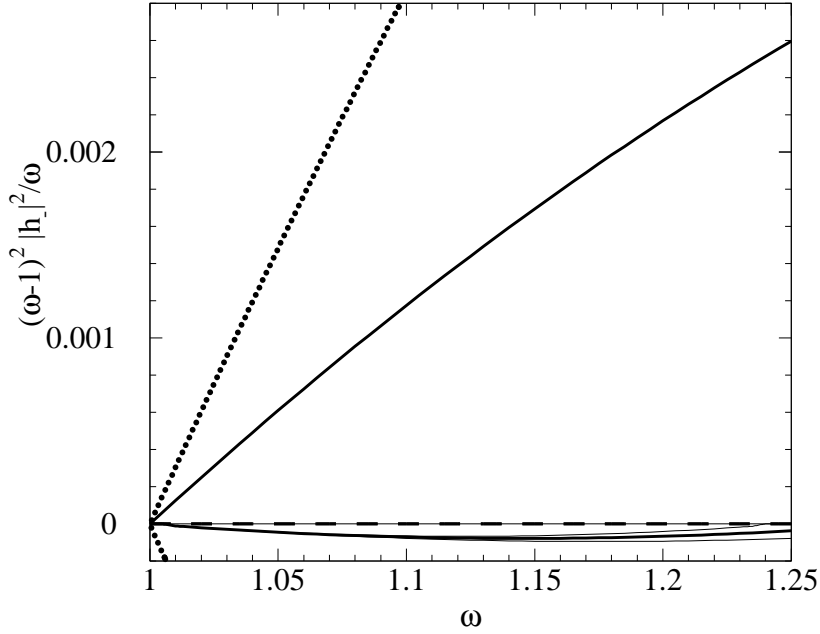} }
\tighten{
\caption[]{\it Upper and lower bounds on $(\omega-1)^2
|h_-(\omega)|^2/\omega$.  The thick solid (dotted) curves are the
bounds with perturbative corrections for HQET parameter set (A), and
$\Delta = 1{\rm\ GeV}\ (2{\rm\ GeV})$.  The dashed curves are the
bounds without perturbative corrections, also for HQET parameter set
(A).  The thin solid curves show the dependence on $\bar\Lambda$ and
$\lambda_1$, using parameter sets (B) and (C), with $\Delta = 1{\rm\
GeV}$.}}
\end{figure}
Notice that in this case, both the upper and lower bounds at tree
level are identically zero.  The perturbative corrections push both
bounds away from zero.  We still plot these negative lower bounding
curves for reference even though the real lower bounds should be zero
since the factors we are bounding are all positive-definite.  Using
$\Delta=2{\rm\ GeV}$ in the calculation widens the upper bound by more
than a factor of 2.

Similarly, the upper bound for $(\omega^2-1) \left| \omega
h_{A_2}(\omega) + h_{A_3}(\omega) \right|^2/(4\omega) \leq
h_{A_3}^{upper}$, where
\begin{eqnarray}
\label{hA3upbound}
h_{A_3}^{upper} &=& - I_1^{(0)AA} + \omega^2 I_2^{(0)AA}
   + ( M_B \omega - M_{D^*})^2 I_4^{(0)AA} 
   - 2\omega (M_B \omega - M_{D^*}) I_4^{(1)AA} \\
&& + \omega^2 I_4^{(2)AA} + 2\omega (M_B \omega - M_{D^*}) I_5^{(0)AA}
  - 2 \omega^2 I_5^{(1)AA} \nonumber \\
&&  - A_1^{(0)AA} + \omega^2 A_2^{(0)AA} 
  + (M_B \omega - M_{D^*})^2 A_4^{(0)AA}
  - 2\omega (M_B \omega - M_{D^*}) A_4^{(1)AA} \nonumber \\
&& + \omega^2 A_4^{(2)AA} + 2\omega (M_B \omega - M_{D^*}) A_5^{(0)AA} 
  - 2 \omega^2 A_5^{(1)AA} ,\nonumber
\end{eqnarray}
while the lower bound $(\omega^2-1) \left| \omega h_{A_2}(\omega) +
h_{A_3}(\omega) \right|^2/(4\omega) \leq h_{A_3}^{lower}$, where
\begin{eqnarray}
\label{hA3lowbound}
h_{A_3}^{lower}&=&h_{A_3}^{upper} - \frac1{M_{D_1}-M_{D^*}} \left\{
  - I_1^{(1)AA} + \omega^2 I_2^{(1)AA}
   + ( M_B \omega - M_{D^*})^2 I_4^{(1)AA} \right. \\
&& \quad - 2\omega (M_B \omega - M_{D^*}) I_4^{(2)AA} + \omega^2 I_4^{(3)AA}
  + 2\omega (M_B \omega - M_{D^*}) I_5^{(1)AA}
  - 2 \omega^2 I_5^{(2)AA} \nonumber \\
&&  \quad - A_1^{(1)AA} + \omega^2 A_2^{(1)AA} 
  + (M_B \omega - M_{D^*})^2 A_4^{(1)AA}
  - 2\omega (M_B \omega - M_{D^*}) A_4^{(2)AA} \nonumber \\
&& \quad 
  +\left.  \omega^2 A_4^{(3)AA} + 2\omega (M_B \omega - M_{D^*}) A_5^{(1)AA} 
  - 2 \omega^2 A_5^{(2)AA} \right\},\nonumber
\end{eqnarray}
Fig.~5 shows these bounds.  
\begin{figure}[t]
\centerline{\epsfysize=11truecm  \epsfbox{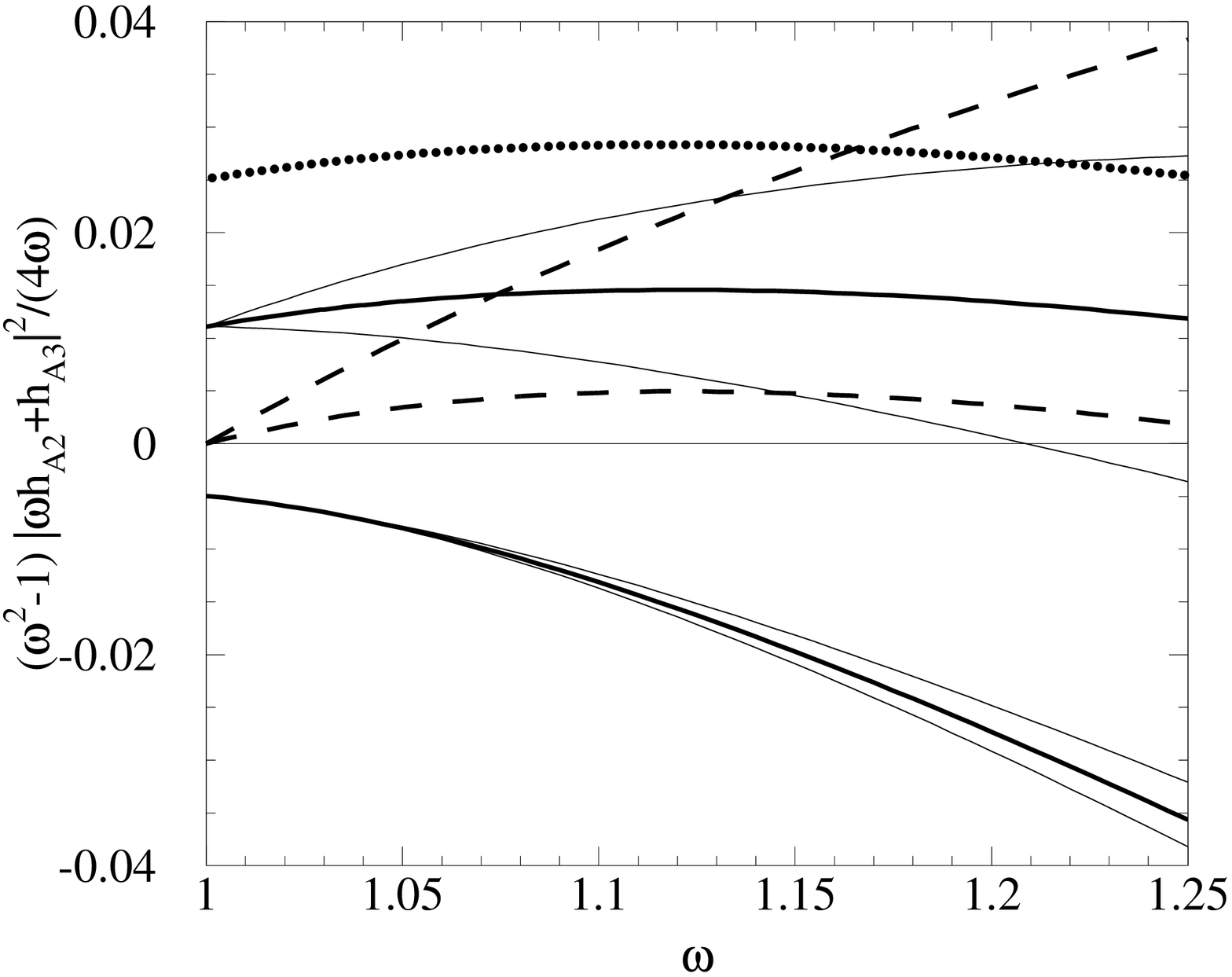} }
\tighten{
\caption[]{\it Upper and lower bounds on $(\omega^2-1) \left| \omega
h_{A_2}(\omega) + h_{A_3}(\omega) \right|^2/(4\omega)$.  The thick
solid (dotted) curves are the bounds with perturbative corrections for
HQET parameter set (A), and $\Delta = 1{\rm\ GeV}\ (2{\rm\ GeV})$.  The
dashed curves are the bounds without perturbative corrections, also
for HQET parameter set (A).  The thin solid curves show the dependence
on $\bar\Lambda$ and $\lambda_1$, using parameter sets (B) and (C), with
$\Delta = 1{\rm\ GeV}$.}}
\end{figure}
The perturbative corrections drive the lower bound negative.  Again,
these lower bounding curves are plotted for reference purposes only.
(The lower bound with $\Delta=2{\rm\ GeV}$ is below the range plotted
in this graph.)  

At ${\mathcal O}(\Lambda_{\rm QCD}/m_Q)$, the upper bounds will not
depend upon $\lambda_1$ and $\lambda_2$.  However, since
$1/(M_{D1}-M_D) \sim 1/\Lambda_{\rm QCD}$ they will affect the lower
bounds.  It is possible to obtain the upper bounds to order ${\mathcal
O}(\Lambda_{\rm QCD}^2/m_Q^2)$, since at this order there are no new
parameters in the OPE.  These corrections only slightly modify the above
upper bounds, so we do not show them here.


\section{Comparison with models}

We choose from the literature the following commonly used form factor
models for comparison with our bounds:
\begin{verse}
(1) ISGW2 Model \cite{ISGW89,SI95}, \\
(2) Light-Front Model \cite{J90,CCH97}, \\
(3) BSW II Model \cite{WSB85,BSW87,KS88}, \\
(4) NS Model \cite{NS92}, \\
(5) COQM Model \cite{MGKIIO99}. \\
\end{verse}

The ISGW2 model is the updated version of the original ISGW model
which incorporates the constraints of heavy quark symmetry and
relativistic corrections.  The BSW II model, derived from the BSW I
model by improving the pole structure of form factors, is considered
with updated pole masses \cite{ALV99}.  The NS model was proposed as a
simplified alternative to the NRSX model \cite{NRSX92} which specifies
the Isgur-Wise function by several strong assumptions such as the pole
structure and the condition for the derivative of the single-pole form
factor.  The form factors are then obtained by employing the heavy
quark symmetry relations.  The COQM model uses a covariant oscillator
quark model to calculate the Isgur-Wise function, and then uses heavy
quark symmetry to relate it to the form factors.

Figs.~6-10 are plots for the different form factors from the models
and the bounding curves.  In plotting these figures, we used $m_b =
4.8 {\rm\ GeV}$, $m_c = 1.4 {\rm\ GeV}$, $\alpha_s = 0.3$
(corresponding to a scale of about $2 {\rm\ GeV}$), $\lambda_2 = 0.12
{\rm\ GeV}^2$, $\Delta = 1 {\rm\ GeV}$ and current PDG data on heavy
meson masses.  As mentioned above, while $\lambda_2$ can be measured
from the mass difference $M_{D^*}-M_D$ to a high accuracy, $\lambda_1$
and $\bar\Lambda$ are not easy to obtain experimentally.  Here we pick
the parameter set (A) discussed above, $\bar\Lambda = 0.4 {\rm\ GeV}$,
$\lambda_1 = -0.2 {\rm\ GeV}^2$.  This uncertainty on $\lambda_1$ and
$\bar\Lambda$ will slightly modify our bounds to the order and
kinematic range we are working, as seen from Figs.~1-5.

Fig.~6 shows the model values of $(\omega+1)^2 \left|h_{A_1}(\omega)
\right|^2/(4\omega)$ along with the corresponding bounds for
comparison.  
\begin{figure}[t]
\centerline{\epsfysize=11truecm  \epsfbox{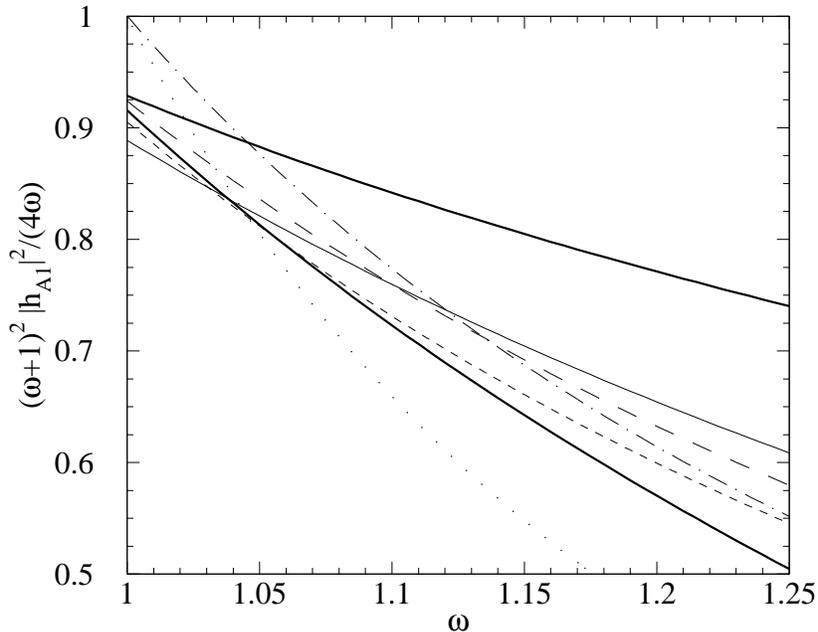} }
\tighten{
\caption[]{\it The model values of $(\omega+1)^2 \left|h_{A_1}(\omega)
\right|^2/(4\omega)$ along with the corresponding bounds for
comparison.  The thick solid lines are the upper and lower bounds.
The thin solid curve is the ISGW2 model.  The long dashed curve is the
LF model, and the short dashed curve is the BSW II model.  The
dot-dashed curve is the NS model.  The dotted curve is the COQM
model.}}
\end{figure}
The first three models have curves falling within or close to the
perturbative bounds, while the NS and COQM models have curves slightly
off the bounds near zero recoil because they are models designed for
the Isgur-Wise function without taking into account the perturbative
corrections.  The COQM model, however, has a curve which falls far
below the lower bound at large $\omega$.

In Fig.~7, the ISGW2 and NS models lie above the upper bound for
$(\omega^2-1)\left|\omega h_{A_2} + h_{A_3} \right|^2/(4\omega)$ at
large recoil, while the other models are closer to our bounds.  For
$\Delta=1{\rm\ GeV}$ they are above our bounds for most of the
kinematic range.  If we take $\Delta = 2{\rm\ GeV}$, however, they
would be within our bounds.  (ISGW2 and NS would still be too large.)
\begin{figure}[t]
\centerline{\epsfysize=11truecm  \epsfbox{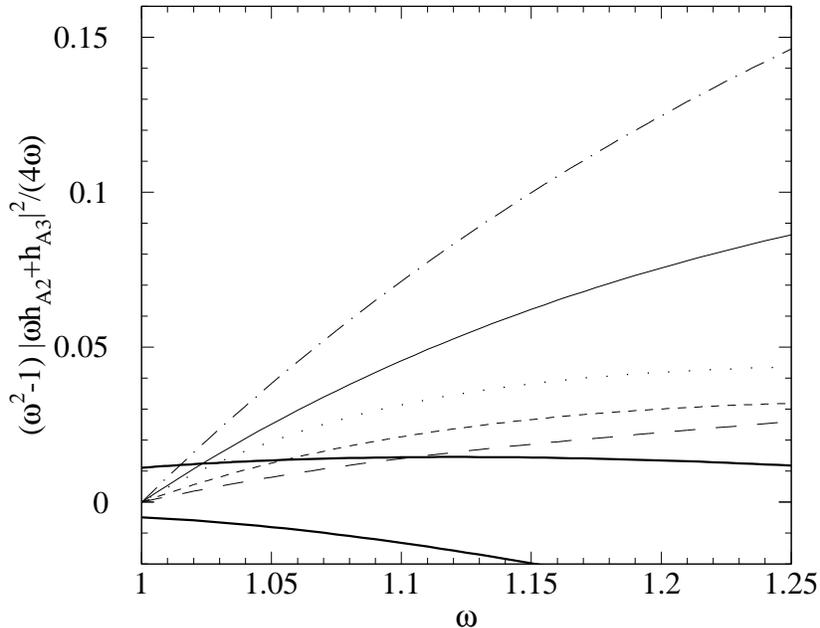} }
\tighten{
\caption[]{\it The model values of $(\omega^2-1)\left|\omega h_{A_2} + h_{A_3}
\right|^2/(4\omega)$ along with the corresponding bounds for
comparison.  The curves are labeled the same as in Fig.~6.}}
\end{figure}

Only the ISGW2 model predicts a curve near our bounds for $(\omega^2
-1)\left| h_V \right|^2$, shown in Fig.~8; the rest are too small.
\begin{figure}[t]
\centerline{\epsfysize=11truecm  \epsfbox{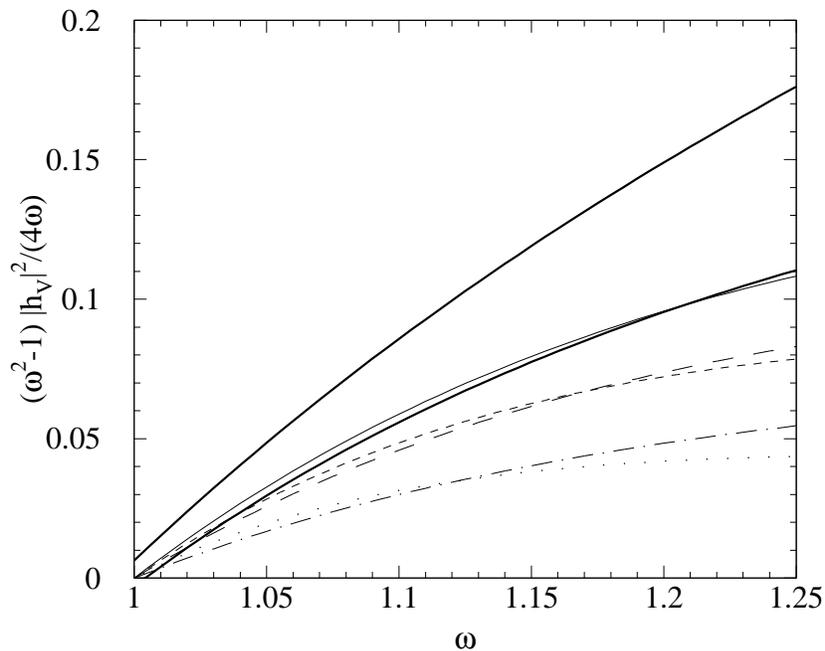} }
\tighten{
\caption[]{\it The model values of $(\omega^2 -1)\left| h_V \right|^2$
along with the corresponding bounds for comparison.  The curves are
labeled the same as in Fig.~6.}}
\end{figure}
The Light-Front and BSW II models will be between our bounds when the
scale is set at $\Delta = 2{\rm\ GeV}$ , but the NS and COQM models
still fall below the lower bound.  

As shown in Fig.~9, the ISGW2 model agrees with our bounds for
$(\omega+1)^2\left| h_+ \right|^2/\omega$ very well.  For the same
reason as in Fig.~6, the NS and COQM models start from the position
predicted by heavy quark symmetry at zero recoil, without perturbative
corrections.  The Light-Front, NS, and COQM models stay below the
lower bound, and the BSW II model starts above the bounds at
$\omega=1$, then cross the bounds and enter the region under the lower
bound.
\begin{figure}[t]
\centerline{\epsfysize=11truecm  \epsfbox{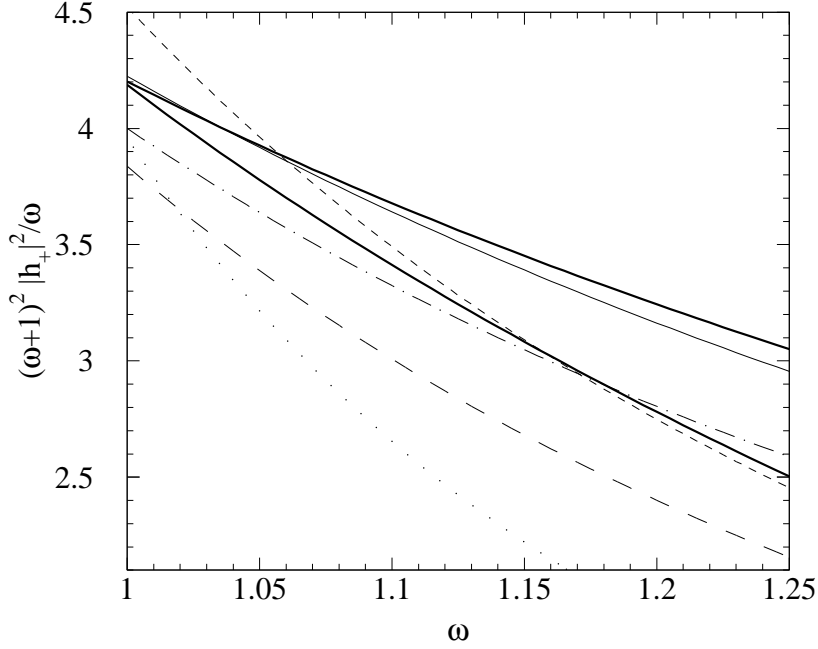} }
\tighten{
\caption[]{\it The model values of $(\omega+1)^2\left| h_+
\right|^2/\omega$ along with the corresponding bounds for comparison.
The curves are labeled the same as in Fig.~6.}}
\end{figure}
At large recoil, the ISGW2, BSW II, and NS model predictions would lie
within the bounds for most of the kinematic range with a larger value
for $\Delta$.

Fig.~10 shows the curves for $(\omega -1)^2\left| h_-
\right|^2/\omega$, where most models are consistent with our bounds
within the kinematic range of interest.  The curve for the COQM model
is zero, since it uses heavy quark symmetry, which gives $h_-(w) = 0$
in the infinite mass limit.
\begin{figure}[t]
\centerline{\epsfysize=11truecm  \epsfbox{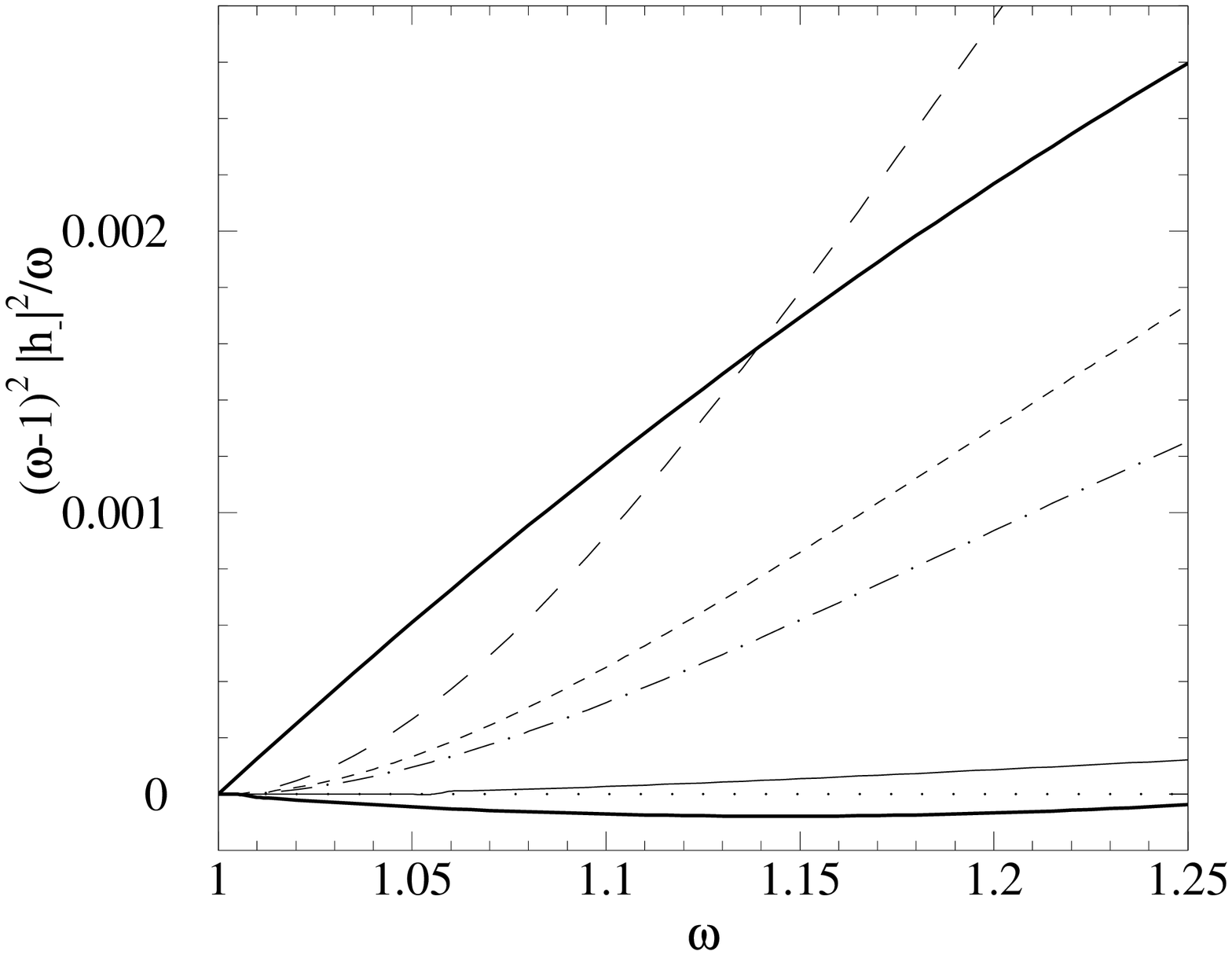} }
\tighten{
\caption[]{\it  The model values of $(\omega -1)^2\left| h_- \right|^2/\omega$
along with the corresponding bounds for comparison.  The curves are
labeled the same as in Fig.~6.}}
\end{figure}

The scale we choose to plot these diagrams is $\Delta = 1 {\rm\ GeV}$,
since this gives the tightest bounds.  Had we chosen $2 {\rm\ GeV}$ as
our working scale, the bounds would be much less stringent and thus
would accommodate the models which originally fell slightly outside
our bounds in these diagrams.


\section{Order $\alpha_s^2 \beta_0$ corrections at zero recoil}

By using the techniques introduced in Ref.~\cite{SV94}, we can
calculate the $\alpha_s^2 \beta_0$ contribution to the structure
functions.  Typically, these corrections are about 90\% of the full
$\alpha_s^2$ rate, so they can be used as a rough estimate of the next
order corrections.  The $O(\alpha_s^2 \beta_0)$ correction to the
structure functions, $T_i^{\alpha_s^2\beta_0}$, can be related to the
$O(\alpha_s)$ correction, $T_i^{\alpha_s}(\lambda)$, calculated with a
gluon mass $\lambda$, as \cite{SV94}
\begin{equation}
T_i^{\alpha_s^2 \beta_0} = -\beta_0\frac{\alpha_s^{(V)}(\Delta)}{4\pi}
\int_0^\infty \frac{d\lambda^2}{\lambda^2} \left(T_i^{\alpha_s}(\lambda)
  - \frac{\Delta^2}{\lambda^2+\Delta^2}T_i^{\alpha_s}(0)\right),
\end{equation}
where $\alpha_s^{(V)}$ is the strong coupling constant evaluated in
the ``V-scheme'' \cite{BLM83}, and is related to the $\overline{\rm
MS}$ coupling constant $\overline{\alpha}_s$ by
\begin{equation}
\alpha_s^{(V)}(\mu) = \overline{\alpha}_s(\mu) + 
   \frac53\frac{\overline{\alpha}_s(\mu)^2}{4\pi}\beta_0 + \cdots .
\end{equation}
We numerically calculate these corrections at zero recoil.  In analogy
to Eq.~(\ref{SFmoments}), we write the corrections to the moments of the
structure as
\begin{equation}\label{SFmoments2}
\frac1{2\pi i}\int d\epsilon\,\epsilon^n\,T_i =  
I^{(n)}_i + A^{(n)}_i,
\end{equation}
where $A^{(n)}_i$ now contains both the $O(\alpha_s)$ and the
$O(\alpha_s^2 \beta_0)$ contribution to the $n^{th}$ moment of the
$i^{th}$ structure function at zero recoil.  The results for the
$A^{(n)}_i$'s for the zeroth and first moments are presented in
Table~1, for $\Delta = 1,2 {\rm\ GeV}$.  The $O(\alpha_s^2 \beta_0)$
contributions are often large, especially for $\Delta = 1 {\rm\
GeV}$.  This may be an indication that we need to go to larger
$\Delta$ to get a reliable perturbative expansion.

\begin{table}[ht]
\begin{tabular}{c|ccc|ccc}  
&  \multicolumn{2}{c}{$\Delta = 1\,$GeV}  & 
  &  \multicolumn{2}{c}{$\Delta = 2\,$GeV} &  \\
& $O(\alpha_s)$ & $O(\alpha_s^2\beta_0)$ &
   & $O(\alpha_s)$ & $O(\alpha_s^2\beta_0)$ & \\ \hline 
$A_1^{(0)AA}$  &  -0.071  &  0.016 && -0.059   & 0.005 & \\
$A_1^{(1)AA}$  &  0.0053  &  0.0040 && 0.024  & 0.018 & \\
$A_2^{(0)AA}$  &  -0.13  &  0.21 &&  -0.042  & 0.22  & \\
$A_2^{(1)AA}$  &  0.086  &  0.057  &&  0.21  & 0.15 & \\
$A_4^{(0)AA}$  &  -0.015  &  0.003  &&  -0.013  & -0.0004 & \\
$A_4^{(1)AA}$  &  0.0023  & 0.0014  &&  0.0054  & 0.0036 & \\
$A_5^{(0)AA}$  &  0.032  &  -0.033 &&  0.015  & -0.034  & \\
$A_5^{(1)AA}$  &  -0.017  &  -0.011 &&  -0.041  & -0.029 & \\
\hline
$A_1^{(0)VV}$  &  0.0063  & 0.0042 && 0.015   & 0.011 & \\
$A_1^{(1)VV}$  &  0.0038  & 0.0022 && 0.017   & 0.011 & \\
$A_2^{(0)VV}$  &  0.26  & 0.13 && 0.32  & 0.24 & \\
$A_2^{(1)VV}$  &  0.062  & 0.042 &&  0.14   & 0.10 & \\
$A_4^{(0)VV}$  &  -0.0060  & 0.0012 &&  -0.0048  & 0.0002 & \\
$A_4^{(1)VV}$  &  0.0014  & 0.0009 && 0.0032   & 0.0022 & \\
$A_5^{(0)VV}$  &  -0.022  & -0.024 && -0.034   & -0.037 & \\
$A_5^{(1)VV}$  &  -0.012  & -0.008 &&  -0.029  & -0.021 & \\
\end{tabular} \vspace{6pt}
\caption{Order $\alpha_s$ and $\alpha_s^2\beta_0$ corrections to the
moments of the structure functions at zero recoil, evaluated for
$\Delta = 1,2{\rm\ GeV}$.  For this table, $\alpha_s = 0.3$, and $n_f
= 4$.}
\end{table}

If we look at the actual combination of structure functions that
appear in our bounds, the situation is more promising.  The
$O(\alpha_s^2 \beta_0)$ corrections to the upper and lower bounds on
the form factors at zero recoil are shown in Table~2 for $\Delta =
1{\rm\ GeV}$.  In this case, the $O(\alpha_s^2 \beta_0)$ corrections
tend to be rather small, so it seems that the perturbative expansion for
the bounds is under control.

\begin{table}[ht]
\begin{tabular}{c|cccc}  
& Tree Level & $O(\alpha_s)$ & $O(\alpha_s^2\beta_0)$ \\ \hline
Upper $\frac{(\omega+1)^2}{4\omega} |h_{A_1}|^2$ &  1 & -0.071 & 0.016 \\
Lower $\frac{(\omega+1)^2}{4\omega} |h_{A_1}|^2$ &  1 & -0.084 & 0.006 \\
\hline
Upper $\frac{\omega^2-1}{4\omega} \left| \omega h_{A_2} + h_{A_3}
\right|^2$ &  0 & 0.011 & 0.007 \\
Lower $\frac{\omega^2-1}{4\omega} \left| \omega h_{A_2} + h_{A_3}
\right|^2$ &  0 & -0.005 & -0.001 \\
\hline
Upper $\frac{\omega^2-1}{4\omega} |h_{V}|^2$ &  0 & 0.006 & 0.004 \\ 
Lower $\frac{\omega^2-1}{4\omega} |h_{V}|^2$ &  0 & -0.003 & -0.001 \\ 
\hline
Upper $\frac{(\omega+1)^2}{\omega} |h_+|^2$ &  4 & 0.20 & -0.05 \\
Lower $\frac{(\omega+1)^2}{\omega} |h_+|^2$ &  4 & 0.19 & -0.06 \\
\hline
Upper $\frac{(\omega-1)^2}{\omega} |h_-|^2$ &  0 & 0 & 0 \\
Lower $\frac{(\omega-1)^2}{\omega} |h_-|^2$ &  0 & 0 & 0
\end{tabular} \vspace{6pt}
\caption{Bound on form factors at zero recoil, evaluated with
$\Delta = 1{\rm\ GeV}$.}
\end{table}

\section{Discussion}

As discussed earlier, our bounds on the form factors are derived to
order $\Lambda_{\rm QCD}/m_Q$ and contain leading perturbative QCD
corrections.  They are constructed to be model independent and can be
used to test various model predictions for the form factors.
Furthermore, they may be used to bound inclusive or exclusive decay
rates.  The validity of the lower bounds requires the extra assumption
of negligible multiparticle production in the decays.  This assumption
is rather mild, since it is true in the large $N_c$ limit and is
supported by current experimental data.  The upper bound is rigorously
valid without this extra assumption.  Therefore, any significant
deviation of the phenomenological form factor beyond the upper bound
indicates the need for some modification of the model.  In general,
our bounds are stricter and more accurate near zero recoil.

The bounds become much less stringent when $\Delta$ is increased.
Therefore, we should use the smallest value of $\Delta$ for which
there is a reasonable perturbative expansion.  We also observed that
the $O(\alpha_s^2 \beta_0)$ corrections to the bounds at zero recoil
are small for $\Delta = 1{\rm\ GeV}$, which in turn suggests that a
perturbative expansion in this problem seems to work well, provided
that the $O(\alpha_s^2 \beta_0)$ corrections dominate in the complete
$O(\alpha_s^2)$ corrections.

Among the models we considered in this work, the ISGW2, LF, and BSW II
models ``pass'' most of the qualifications with $\Delta = 2{\rm\
GeV}$.  For smaller values of $\Delta$, ISGW2 seems to do the best.
However, all models fail, over essentially the whole kinematic range,
for the form factor combination $\frac{\omega^2-1}{4\omega} \left|
\omega h_{A_2} + h_{A_3} \right|^2$, and thus the models should be
modified to satisfy this bound.


\acknowledgments

The authors would like to thank Fred Gilman and Ira Rothstein for
useful discussions.  This work was supported in part by the Department
of Energy under Grant No. DE-FG02-91ER40682. 


\begin{appendix}
\section*{\ Radiative corrections to the structure functions}

%
%
%
%

The structure functions from perturbative corrections are written in
the form $T_i^{JJ} = (\alpha_s/\pi)[U_i^{JJ} + \left( \omega-1 \right)
V_i^{JJ}]$ for the $i$th structure function and current $J$.  $m_b$
is the mass of the $b$-quark, $m_c$ is the mass of the $c$-quark, $z
\equiv m_c/m_b$, $\omega = v \cdot v^{\prime}$, and $\lambda$ is the
mass of the gluon used for infrared regularization.  We have taken
$\lambda \to 0$ wherever possible.

\begin{eqnarray}
U_1^{VV} &=&
 \frac{\epsilon\,\left( \epsilon + 2\,m_b\,z \right) \, \left[ 2\,\epsilon^2
         + 4\,m_b\,\epsilon\,z + m_b^2\,\left( 3 - 2\,z + 3\,z^2
         \right)  \right] }
      {9\, m_b^2\,\left( \epsilon + m_b\,z \right)^3}, \\
\nonumber\\
V_1^{VV} &=&
-\left[ 
   \frac43 
   + \frac{ 1 + z }{1 - z} \log(z)
\right] \delta (\epsilon) \nonumber \\
&&
+ \frac{z}{45\, m_b\,{{\left( \epsilon + m_b\,z \right) }^5}\,
     \left( \epsilon + 2\,m_b\,z \right) }
 \left[ 
    10\,\epsilon^6 + 58\,m_b\,\epsilon^5\,z
    - m_b^2\,\epsilon^4\,\left( 15 + 18\,z - 133\,z^2 \right)
     \right. \nonumber \\
&& \qquad
    + m_b^3\,\epsilon^3\,z\,\left(35 - 94\,z + 151\,z^2 \right)
    + 2\,m_b^4\,\epsilon^2\,z^2\,
       \left( 135 - 110\,z + 47\,z^2 \right)
       \nonumber \\
&& \qquad 
 \left.
    + 8\,m_b^5\,\epsilon\,z^3\,\left( 45 - 31\,z + 5\,z^2 \right)
    + 80\,m_b^6\,\left( 1 - z \right) \,z^4
 \right] , \\
\nonumber \\
U_2^{VV} &=&
-\left[
     \frac{8}{3\,z} 
   + \frac{4\,\left( 1 + z \right) }{3\,\left( 1 - z \right) \,z}
      \log (z)
\right] \delta (\epsilon) \nonumber \\
&&
- \frac{2}{45\, m_b^2\,\left( \epsilon + m_b\,z \right)^5}
 \left[ 
      \epsilon^6
    - 3\,m_b\,\epsilon^5\,\left( 1 - 2\,z \right)
    + m_b^2\,\epsilon^4\,\left( 28 - 13\,z + 13\,z^2 \right) 
 \right. \nonumber \\
&& \qquad
    - 2\,m_b^3\,\epsilon^3\,
        \left( 23 - 59\,z + 11\,z^2 - 6\,z^3 \right)
    - 2\,m_b^4\,\epsilon^2\,z\, 
       \left( 63 - 96\,z + 10\,z^2 - 2\,z^3 \right) 
       \nonumber \\
&& \qquad
 \left.
    - 4\,m_b^5\,\epsilon\,z^2\,\left( 27 - 34\,z + 2\,z^2 \right) 
    - 40\,m_b^6\,\left( 1 - z \right) \,z^3
 \right] , \\
\nonumber \\
V_2^{VV} &=&
\left[
     \frac{2\,\left( 19 - 62\,z + 19\,z^2 \right) }
           {27\,\left( 1 - z \right)^2\,z}
   - \frac{4\,\left( 1 - 3\,z + 9\,z^2 - 3\,z^3 \right) }
           {9\,\left( 1 - z \right)^3\,z} \log (z)
   + \frac{16}{9\,z} \log \left(\frac{\lambda}{m_b}\right)
\right] \delta (\epsilon) \nonumber \\
&&
- \frac{2}{315\, m_b\,z\,\left( \epsilon + m_b\,z \right) ^7\,
              \left( \epsilon + 2\,m_b\,z \right) }
  \left[ 
     7\,\epsilon^8\,z^2
   + \epsilon^7\,m_b\,\left( 280 + 47\,z^3 \right) 
  \right. \nonumber \\
&& \qquad
   + \epsilon^6\,m_b^2\,z\,
         \left( 2520 - 196\,z - 26\,z^2 + 131\,z^3 \right) 
         \nonumber \\
&& \qquad
   + \epsilon^5\,m_b^3\,z^2\,
         \left( 10444 - 1216\,z - 186\,z^2 + 209\,z^3 \right) 
         \nonumber \\
&& \qquad
   + 2\,\epsilon^4\,{m_b^4}\,z^3\,
        \left( 12060 - 1537\,z - 233\,z^2 + 109\,z^3 \right) 
        \nonumber \\
&& \qquad
   + 2\,\epsilon^3\,m_b^5\,z^4\,
        \left( 15960 - 1905\,z - 277\,z^2 + 70\,z^3 \right) 
        \nonumber \\
&& \qquad
   + 4\,\epsilon^2\,m_b^6\,z^5\,
        \left( 6061 - 550\,z - 96\,z^2 + 10\,z^3 \right) 
   + 8\,\epsilon\,m_b^7\,z^6\,
        \left( 1306 - 64\,z - 17\,z^2 \right) 
        \nonumber \\
&& \qquad
  \left.
   + 56\,m_b^8\,z^7\,\left( 33 + 2\,z \right) 
  \right] 
+ \frac{16\,\sqrt{\epsilon^2 - \lambda^2}}
       {9\,\epsilon^2\,z}
+ \frac{8\,\lambda^2\,\sqrt{\epsilon^2 - \lambda^2}}
       {9\,\epsilon^4\,z}
, \\
\nonumber \\
U_4^{VV} &=&
\left[
     \frac{2}{3\,m_b^2\,\left( 1 - z \right) \,z}
   + \frac{2}{3\,m_b^2\,\left( 1 - z \right)^2\,z} \log (z)
\right] \delta (\epsilon) \\
&&
+ \frac{  2\,\left[ 4\,\epsilon^4 
       + m_b\,\epsilon^3\,\left( 11 + 16\,z \right) 
       + m_b^2\,\epsilon^2\,z\,\left( 21 + 26\,z \right) 
       + 2\,m_b^3\,\epsilon\,z^2\,\left( 9 + 10\,z \right) 
       + 20\,m_b^4\,z^3 \right] }
      {45\, m_b^2\,\left( \epsilon + m_b\,z \right)^5}, \nonumber \\
\nonumber \\
V_4^{VV} &=&
-\left[
     \frac{2\,\left( 2 - z + 5\,z^2 \right) }
           {9\,m_b^2\,\left( 1 - z \right)^3\,z}
   + {\frac{2\,\left( 1 - 2\,z + 3\,z^2 \right) }
           {3\,m_b^2\,\left( 1 - z \right)^4\,z}} \log (z)
\right] \delta (\epsilon) \nonumber \\
&&
- \frac{4\,z}{315\, m_b\,\left( \epsilon + m_b\,z \right)^7}
 \left[
      14\,\epsilon^5
    + m_b\,\epsilon^4\,\left( 77 + 62\,z \right)
    + m_b^2\,\epsilon^3\,z\,\left( 125 + 129\,z \right) 
 \right. \nonumber \\
&& \qquad 
 \left.
    + m_b^3\,\epsilon^2\,z^2\,\left( 135 + 143\,z \right) 
    + m_b^4\,\epsilon\,\left( 323 - 28\,z \right) \, z^3
    + 56\,m_b^5\,z^4 
 \right] , \\
\nonumber \\
U_5^{VV} &=&
\left[
     \frac{3 - 5\,z}{3\,m_b\,\left( 1 - z \right) \,z}
   + \frac{ 1 - 3\,z^2 }
           {3\,m_b\,\left( 1 - z \right)^2\,z} \log (z)
\right] \delta (\epsilon) \nonumber \\
&&
- \frac{1}{45\, m_b^2\,\left( \epsilon + m_b\,z \right)^5}
 \left[
    2\,\epsilon^5
    - 2\,m_b\,\epsilon^4\,\left( 2 - 5\,z \right) 
    + m_b^2\,\epsilon^3\,\left( 57 - 22\,z + 21\,z^2 \right) 
 \right. \nonumber \\
&& \qquad
    + m_b^3\,\epsilon^2\,z\,\left( 147 - 46\,z + 23\,z^2 \right) 
    + 2\,m_b^4\,\epsilon\,z^2\,\left( 63 - 18\,z + 5\,z^2 \right) 
    \nonumber \\
&& \qquad
 \left.
    + 20\,m_b^5\,\left( 3 - z \right) \,z^3
 \right], \\
\nonumber \\
V_5^{VV} &=&
\left[
     - \frac{ 13 - 93\,z + 69\,z^2 - 25\,z^3 }
           {27\,m_b\,\left( 1 - z \right)^3\,z}
   + \frac{ 5 - 14\,z + 42\,z^2 - 30\,z^3 + 9\,z^4 }
           {9\,m_b\,\left( 1 - z \right)^4\,z} \log (z)
\right. \nonumber \\
&& \qquad
\left.
   - \frac{8}{9\,m_b\,z} \log \left (\frac{\lambda}{m_b}\right)
\right] \delta (\epsilon) 
- \frac{8\,\sqrt{\epsilon^2 - \lambda^2}}
       {9\,\epsilon^2\,m_b\,z}
- \frac{4\,\lambda^2\,\sqrt{\epsilon^2 - \lambda^2}}
       {9\,\epsilon^4\,m_b\,z} \nonumber \\
&&
+ \frac{2}{315\,m_b\,z\,\left( \epsilon + m_b\,z \right)^7\,
              \left( \epsilon + 2\,m_b\,z \right) }
  \left[ 
     140\,\epsilon^7
   + 2\,\epsilon^6\,m_b\,z\,
        \left( 630 - 7\,z + z^2 \right) 
  \right. \nonumber \\
&& \qquad
   + 7\,\epsilon^5\,m_b^2\,z^2\,
        \left( 757 - 16\,z + 3\,z^2 \right) 
   + \epsilon^4\,{m_b^3}\,z^3\,
        \left( 12339 - 388\,z + 85\,z^2 \right) 
        \nonumber \\
&& \qquad
   + \epsilon^3\,m_b^4\,z^4\,
         \left( 16345 - 580\,z + 131\,z^2 \right) 
   + \epsilon^2\,m_b^5\,z^5\,
        \left( 12715 - 436\,z + 51\,z^2 \right) 
        \nonumber \\
&& \qquad
  \left.
   + 2\,\epsilon\,m_b^6\,z^6\,
        \left( 2963 - 149\,z - 7\,z^2 \right) 
   + 28\,m_b^7\,z^7\,\left( 37 + z \right) 
  \right].
\end{eqnarray}
\begin{eqnarray}
U_1^{AA} &=&
-\left[
    \frac{16}{3} 
   + \frac{2\,\left( 1 + z \right) }{1 - z} \log (z)
\right] \delta (\epsilon) \nonumber \\
&&
+ \frac{\epsilon\,\left( \epsilon + 2\,m_b\,z \right) \, \left( 2\,\epsilon^2
       + 4\,\epsilon\,m_b\,z + m_b^2\,
          \left( 3 + 2\,z + 3\,z^2 \right)  \right) }
      {9\, m_b^2\,\left( \epsilon + m_b\,z \right)^3}, \\
\nonumber \\
V_1^{AA} &=&
\left[
     \frac{4\,\left( 11 - 52\,z + 11\,z^2 \right) }
           {27\,\left( 1 - z \right)^2}
   - \frac{ 7 - 3\,z + 45\,z^2 - 9\,z^3 }
           {9\,\left( 1 - z \right)^3} \log (z)
   + \frac{16}{9} \log \left(\frac{\lambda}{m_b}\right)
\right]\delta (\epsilon) \nonumber \\
&&
+ \frac{1}{45\,m_b\, \left( \epsilon + m_b\,z \right)^5}
  \left[
     10\,\epsilon^5\,z
   - 2\,\epsilon^4\,m_b\,\left( 40 - 19\,z^2 \right) 
   - \epsilon^3\,m_b^2\,z\,
        \left( 415 + 38\,z - 57\,z^2 \right) 
  \right. \nonumber \\
&& \qquad
   - \epsilon^2\,m_b^3\,z^2\,
        \left( 735 + 110\,z - 37\,z^2 \right) 
   - 4\,\epsilon\,m_b^4\,z^3\,
        \left( 125 + 22\,z - 5\,z^2 \right) 
        \nonumber \\
&& \qquad
  \left.
   - 40\,m_b^5\,z^4\,\left( 4 + z \right) 
  \right] 
+ \frac{16\,\sqrt{\epsilon^2 - \lambda^2}}
       {9\,\epsilon^2} 
+ \frac{8\,\lambda^2\, \sqrt{\epsilon^2 - \lambda^2}}
       {9\,\epsilon^4}, \\
\nonumber \\
U_2^{AA} &=&
-\left[
   \frac4z 
   + \frac{4\,\left( 1 + z \right) }
           {3\,z\,\left( 1 - z \right) } \log (z)
\right] \delta (\epsilon) \nonumber \\
&&
- \frac{2}{45\, m_b^2\,\left( \epsilon + m_b\,z \right)^5}
 \left[ 
      \epsilon^6
    - 3\,\epsilon^5\,m_b\,\left( 1 - 2\,z \right) 
    + \epsilon^4\,m_b^2\, \left( 28 - 17\,z + 13\,z^2 \right) 
 \right. \nonumber \\
&& \qquad
    - 2\,\epsilon^3\,m_b^3\,
        \left( 23 - 53\,z + 19\,z^2 - 6\,z^3 \right) 
    - 2\,\epsilon^2\,m_b^4\,z\,
         \left( 75 - 78\,z + 20\,z^2 - 2\,z^3 \right) \nonumber \\
&& \qquad
 \left.
    - 4\,\epsilon\,m_b^5\,z^2\, \left( 39 - 28\,z + 4\,z^2 \right) 
    - 40\,m_b^6\,\left( 1 - z \right) \,z^3
 \right] , \\
\nonumber \\
V_2^{AA} &=&
\left[
     \frac{2\,\left( 43 - 110\,z + 43\,z^2 \right) }
           {27\,\left( 1 - z \right)^2\,z}
   - \frac{4\,\left( 1 - 3\,z - 9\,z^2 + 3\,z^3 \right) }
           {9\,\left( 1 - z \right)^3\,z} \log (z)
   + \frac{16}{9\,z} \log \left(\frac{\lambda}{m_b}\right)
\right] \delta (\epsilon) \nonumber \\
&&
- \frac{2}{315\, m_b\,z\,\left( \epsilon + m_b\,z \right)^7\,
              \left( \epsilon + 2\,m_b\,z \right) }
  \left[ 
     7\,\epsilon^8\,z^2
   + \epsilon^7\,m_b\,\left( 280 + 47\,z^3 \right) 
  \right. \nonumber \\
&& \qquad
   + \epsilon^6\,m_b^2\,z\,
         \left( 2520 - 196\,z + 2\,z^2 + 131\,z^3 \right) 
         \nonumber \\
&& \qquad
   + \epsilon^5\,m_b^3\,z^2\,
         \left( 10444 - 1048\,z + 18\,z^2 + 209\,z^3 \right) 
         \nonumber \\
&& \qquad
   + 2\,\epsilon^4\,m_b^4\,z^3\,
         \left( 12312 - 1069\,z + 29\,z^2 + 109\,z^3 \right) 
         \nonumber \\
&& \qquad
   + 2\,\epsilon^3\,m_b^5\,z^4\,
         \left( 17028 - 1077\,z + 13\,z^2 + 70\,z^3 \right) 
         \nonumber \\
&& \qquad
   + 4\,\epsilon^2\,m_b^6\,z^5\,
         \left( 6601 - 316\,z - 30\,z^2 + 10\,z^3 \right) 
   + 8\,\epsilon\,m_b^7\,z^6\,
         \left( 1282 - 58\,z - 13\,z^2 \right) 
         \nonumber \\
&& \qquad
  \left.
   + 56\,m_b^8\,z^7\,\left( 33 + 2\,z \right) 
  \right]
+ \frac{16\,\sqrt{\epsilon^2 - \lambda^2}}
       {9\,\epsilon^2\,z}
+ \frac{8\,\lambda^2\,\sqrt{\epsilon^2 - \lambda^2}}
       {9\,\epsilon^4\,z}, \\
\nonumber \\
U_4^{AA} &=&
-\left[
   \frac{2\,\left(1+ 3\,z\right)}{3\,m_b^2\,z\,\left(1-z\right)^2} 
   - \frac{2\,\left( 1 -5\,z \right) }
           {3\,m_b^2\,z\,\left( 1 - z \right)^3 } \log (z)
\right] \delta (\epsilon) \\
&&
 + \frac{2\,\left[ 
         4\,\epsilon^4
       + \epsilon^3\,m_b\,\left( 11 + 16\,z \right) 
       + \epsilon^2\,m_b^2\,z\,\left( 45 + 26\,z \right) 
       + 2\,\epsilon\,m_b^3\,z^2\,\left( 33 + 10\,z \right) 
       + 20\,m_b^4\,z^3
        \right] }
      {45\,m_b^2\, \left( \epsilon + m_b\,z \right)^5},
\nonumber \\
\nonumber \\
V_4^{AA} &=&
\left[
   \frac{2\,\left(6 - 19\,z - 46\,z^2 + 11\,z^3 \right)}
       {9\,m_b^2\,z\,\left(1-z\right)^4} 
   - \frac{2\,\left(1 - 7\,z + 25\,z^2 - 3\,z^3 \right) }
           {3\,m_b^2\,z\,\left( 1 - z \right)^5 } \log (z)
\right] \delta (\epsilon) \nonumber \\
&&
 -  \frac{4\,z}{315\,m_b\, \left( \epsilon + m_b\,z \right)^7}
 \left[ 
      14\,\epsilon^5
    + \epsilon^4\,m_b\,\left( 77 + 62\,z \right) 
    + \epsilon^3\,m_b^2\,z\,\left( 377 + 129\,z \right) 
 \right. \nonumber \\
&& \qquad
 \left.
    + \epsilon^2\,m_b^3\,z^2\,\left( 699 + 143\,z \right) 
    + \epsilon\,m_b^4\,\left( 275 - 28\,z \right) \,z^3
    + 56\,m_b^5\,z^4
 \right] , \\
\nonumber \\
U_5^{AA} &=&
\left[
     \frac{7 - 5\,z}{3\,m_b\,\left( 1 - z \right) \,z}
   + \frac{\left( 1 + 4\,z - 3\,z^2 \right) }
           {3\,m_b\,\left( 1 - z \right)^2\,z} \log (z)
\right] \delta (\epsilon) \nonumber \\
&&
+ \frac{1}{45\,m_b^2\,
     \left( \epsilon + m_b\,z \right)^5}
 \left[ 
    - 2\,\epsilon^5
    + 2\,\epsilon^4\,m_b\,\left( 2 - 5\,z \right) 
    - \epsilon^3\,m_b^2\,
         \left( 57 - 10\,z + 21\,z^2 \right) 
 \right. \nonumber \\
&& \qquad
    - \epsilon^2\,m_b^3\,z\,
         \left( 195 - 10\,z + 23\,z^2 \right) 
    - 2\,\epsilon\,m_b^4\,z^2\,
         \left( 111 - 6\,z + 5\,z^2 \right) 
 \nonumber \\
&& \qquad
 \left.
    - 20\,m_b^5\,\left( 3 - z \right) \,z^3
 \right] , \\
\nonumber \\
V_5^{AA} &=&
-\left[
     \frac{\left( 61 - 225\,z + 81\,z^2 - 25\,z^3 \right) }
           {27\,m_b\,\left( 1 - z \right)^3\,z}
   - \frac{\left( 5 - 26\,z + 90\,z^2 - 42\,z^3 + 9\,z^4 \right) }
           {9\,m_b\,\left( 1 - z \right)^4\,z} \log (z)
\right. \nonumber \\
&& \qquad
\left.
   + \frac{8}{9\,m_b\,z} \log \left(\frac{\lambda}{m_b}\right)
\right] \delta (\epsilon) 
- \frac{8\,\sqrt{\epsilon^2 - \lambda^2}}
       {9\,\epsilon^2\,m_b\,z} 
- \frac{4\,\lambda^2\,\sqrt{\epsilon^2 - \lambda^2}}
       {9\,\epsilon^4\,m_b\,z} \nonumber \\
&&
+  \frac{2}{315\,m_b\,z\, \left( \epsilon + m_b\,z \right)^7\,
              \left( \epsilon + 2\,m_b\,z \right) } 
 \left[ 
    140\,\epsilon^7
  + 2\,\epsilon^6\,m_b\,z\,
       \left( 630 - 7\,z + z^2 \right) 
 \right. \nonumber \\
&& \qquad
  + 7\,\epsilon^5\,m_b^2\,z^2\,
       \left( 757 - 4\,z + 3\,z^2 \right) 
  + \epsilon^4\,m_b^3\,z^3\,
       \left( 12843 + 80\,z + 85\,z^2 \right) 
       \nonumber \\
&& \qquad
  + \epsilon^3\,m_b^4\,z^4\,
       \left( 18481 + 248\,z + 131\,z^2 \right) 
  + \epsilon^2\,m_b^5\,z^5\,
       \left( 14875 + 32\,z + 51\,z^2 \right) 
       \nonumber \\
&& \qquad
 \left.
  + 2\,\epsilon\,m_b^6\,z^6\,
       \left( 2867 - 137\,z - 7\,z^2 \right) 
  + 28\,m_b^7\,z^7\,\left( 37 + z \right) 
 \right]. 
\end{eqnarray}

\end{appendix}


{\tighten

}

\end{document}